  \documentstyle[twocolumn,floats,prb,aps,psfig]{revtex}
\begin{document}
\draft
\title{Comparison of structural transformations and superconductivity in
compressed Sulfur and Selenium}
\author{Sven P. Rudin}
\address{Los Alamos National Laboratory, Los Alamos, NM 87545}
\author{Amy Y. Liu, J. K. Freericks, and A. Quandt}
\address{Department of Physics, Georgetown University, Washington, DC 20057}
\date{\today}
\maketitle
\begin{abstract}
Density-functional calculations are presented for
high-pressure structural phases of S and Se.
The structural phase diagrams, phonon spectra,
electron-phonon coupling, and superconducting properties
of the isovalent elements are compared. We find that
with increasing pressure, Se adopts a sequence
of ever more closely packed structures ($\beta$-Po, bcc, fcc), while
S favors more open structures ($\beta$-Po, simple cubic, bcc).
These differences are shown to be attributable to differences in
the S and Se core states. All the compressed phases of
S and Se considered are calculated to have weak to moderate
electron-phonon coupling strengths consistent with superconducting
transition temperatures in the range of 1 to 20 K.
Our results compare well with experimental
data on the $\beta$-Po$\rightarrow$bcc transition pressure in Se
and on the superconducting transition temperature in 
$\beta$-Po S.
Further experiments are suggested
to search for the other structural phases predicted at
higher pressures and to test theoretical
results on the electron-phonon interaction and
superconducting properties.
\end{abstract}
\pacs{61.50.Ks, 63.20.Kr, 74.25.Kc, 74.62.Fj
\hfill
LA-UR 00-5587}

\section{INTRODUCTION}  %%%%%%%%%%%%%%%%%%%%%%%%%%%%%%%%%%%%%%%

The chalcogens exist in a rich variety of crystal 
structures at ambient pressures and upon 
compression.\cite{akahama93a,luo93,akahama93b,aoki80,parthasarathy88} 
At low pressures, S, Se, and Te crystallize in different 
sequences of insulating two-fold-coordinated molecular or polymeric
structures. With enough compression, however, all three elements
transform into the same body-centered orthorhombic (bco) crystal structure.
The bco structure is comprised of puckered layers with each
site having a coordination number of four. 
With further compression, the 
more three-dimensional six-fold coordinated rhombohedral $\beta$-Po 
structure becomes stable in all three systems. At even higher 
pressures, Se and Te are observed to transform from the $\beta$-Po 
structure to the bcc structure, continuing the trend of increasing 
coordination and closer packing with pressure. Though is has been
speculated that S should also undergo a $\beta$-Po $\rightarrow$ bcc
transition, recent first-principles calculations
predict $\beta$-Po S to become unstable first to the relatively
open simple-cubic (sc) structure.\cite{rudin99} 
The calculations find simple-cubic S to be favored over a large range of 
pressures, 
with the bcc phase eventually becoming stable at very high pressures.

Electrical measurements show that S, Se, and Te are 
all metallic in the bco structure.\cite{metalrefs}
Further, the metallic bco phases of S, Se, and Te are 
observed to be superconducting with transition temperatures
of approximately
10, 5, and 3 K, respectively.\cite{bundy80,akahama92}
Recent magnetic measurements show that the superconducting transition
temperature in S jumps abruptly to 17 K upon transformation to the
$\beta$-Po structural phase.\cite{struzhkin97}
This is among the highest 
transition temperatures observed in elemental solids. 
First-principles calculations indicate that this high transition temperature
results from a
moderate electron-phonon coupling combined with
a large average phonon frequency, which
sets the scale for the transition temperature.\cite{rudin99}
Earlier calculations had predicted the hypothetical bcc
phase of S to be superconducting with a similar transition
temperature of about 15 K near 550 GPa.\cite{zakharov95}
As yet, no measurements of $T_C$ have been reported 
on the higher-pressure phases of the heavier chalcogens.

In this paper, we report
density-functional calculations of
the structural and superconducting transitions in S and Se
in the $\beta$-Po and higher-pressure phases.
Though the two materials are chemically similar, they display
striking differences in their properties under pressure.
In addition to
the simple-cubic structure's viability as a high-pressure phase for S
but not Se, the stability of the fcc structure is another
difference between the materials.
An analysis of the contributions to the
total energy shows that these structural differences
can be understood in terms of differences between the S and Se cores. 
Although some of the trends in the
calculated electron-phonon coupling and superconducting properties are
similar in the two 
materials, the superconducting transition temperature is higher
in S than in Se, even within the same structural phase.
A comparison of the
phonon spectra and electron-phonon coupling in these
materials helps to explain the differences. 

The remainder of this paper is organized as follows. In 
Sec. II, we compare and contrast the
calculated high-pressure structural phase diagrams
for S and Se. In Sec. III, the
calculated vibrational spectra and electron-phonon coupling parameters
for the high-pressure phases 
are presented and discussed.  We also discuss
ways in which our results can be tested and compared
with diamond-anvil-cell-based experiments, including measurements
of superconducting properties, transport coefficients, and optical 
conductivity. Concluding remarks are given in Sec. IV.

\section{HIGH-PRESSURE STRUCTURAL PHASES}  %%%%%%%%%%%%%%%%%%%%%%%%%%%%%%

\subsection{Method}

Zero-temperature structural energetics are calculated
using the plane-wave pseudopotential method within
the local density approximation. The pseudopotentials
are generated using the Troullier-Martins method.\cite{troullier91} 
Kinetic-energy cutoffs of 70 Ry and 40 Ry are used for the
plane-wave expansion of the Kohn-Sham orbitals in S and Se, respectively. 
The correlation potential is approximated using the 
Perdew-Zunger parameterization of the electron-gas results,\cite{perdew81}
and the nonlinearity of the exchange and correlation
interaction between the core and valence charge
densities is treated using a partial core.\cite{louie82}
The Brillouin zone
for each structure is sampled on Monkhorst-Pack meshes of at least
$20^3$ points.\cite{monkhorst76}

\subsection{Results and discussion} 

\begin{figure}
\centerline{\psfig{file=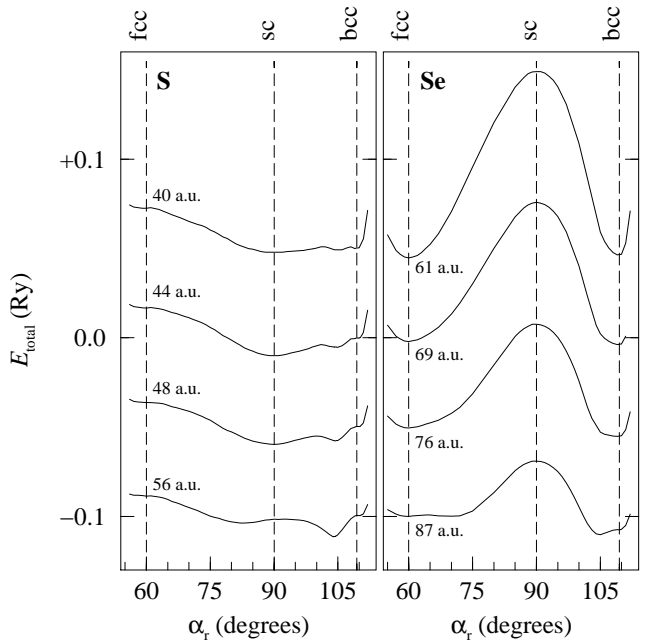,width=3.3in,clip=}}
\caption{
Total energy versus rhombohedral angle $\alpha_r$ at constant volumes for
S and Se.
At moderate compression both elements favor the $\beta$-Po structure
($\alpha_r\approx104^\circ$);
increased compression results in the sequence
$\beta$-Po$\rightarrow$sc$\rightarrow$bcc for S and
$\beta$-Po$\rightarrow$bcc$\rightarrow$fcc for Se.
}
\label{fig1}
\end{figure}

The high-pressure crystal structures considered here for the chalcogens 
can all be described in
terms of a single-atom rhombohedral unit cell with rhombohedral angles
$\alpha_r$ of 60$^\circ$ (fcc), 90$^\circ$ (sc), approximately 104$^\circ$
($\beta$-Po), and 109.47$^\circ$ (bcc).\cite{bcofootnote}
Figure~\ref{fig1} shows the total energy
for S and Se calculated as a function of the rhombohedral angle
at constant unit-cell volumes.
The curves for the two materials are strikingly different, leading to 
different sequences of high-pressure phases. In particular,
for the range of volumes shown in Fig.~\ref{fig1}, the 
simple-cubic structure is never competitive in energy
with the other structures considered for Se, while the fcc
structure is never favorable for S.

For the largest unit-cell volume shown, corresponding
to roughly 80 GPa in Se, the $\beta$-Po structure
with $\alpha_r \approx 104^\circ$ is lowest in energy. At 
a smaller volume of 76 a.u. (roughly 140 GPa),
the local minimum for $\beta$-Po Se has disappeared,
and the bcc structure 
is favored. At even smaller volumes, the minimum at $\alpha_r=60^\circ$
drops below that at 109.5$^\circ$, indicating a bcc to fcc transition.
Comparison of the enthalpies 
$H=E_{\rm total} +pV$
calculated for the different
structures yields a transition pressure of 120 GPa for 
the $\beta$-Po$\rightarrow$bcc transition, which is in reasonable agreement 
with the experimental value of 140 GPa. For the bcc$\rightarrow$fcc transition,
we predict a transition pressure of about 260 GPa.  

For S, the $\beta$-Po structure, which is favored at a unit-cell volume
of 56 a.u., is calculated to become unstable
under compression with respect to the simple-cubic structure,
and the simple-cubic
structure subsequently becomes unstable to the bcc structure (not shown
in Fig.~\ref{fig1}).
Using enthalpies calculated for the static lattices,
we find transition pressures of 260 and 540 GPa for these structural 
transformations. Since we find that the structural phases of S differ 
significantly in their average phonon frequencies (see Sec. III),
it is important to take into account zero-point contributions 
to the energies. 
Inclusion of zero-point energies shifts the estimated 
pressures in S to about 280 GPa for the $\beta$-Po$\rightarrow$sc transition
and about 500 GPa for the sc$\rightarrow$bcc transition.

In terms of coordination number and packing, Se follows
the expected trend of increasing coordination with pressure,
going from six ($\beta$-Po) to eight (bcc) to twelve (fcc). 
With each transformation, Se adopts a more closely packed structure.
Sulfur, on the other hand, transforms from the six-fold coordinated
$\beta$-Po structure to the sc structure, which is also six-fold 
coordinated, but is more open. 
(The ratio of next-nearest to nearest-neighbor distances
is 1.24 and 1.41 in the $\beta$-Po and sc structures, respectively.)
While the more highly coordinated bcc 
structure is adopted at large compressions, the even more densely packed 
fcc structure is never favored. The key to these differences in the 
high-pressure phase diagrams of these isovalent elements lies
in their cores.

\begin{figure}
\centerline{\psfig{file=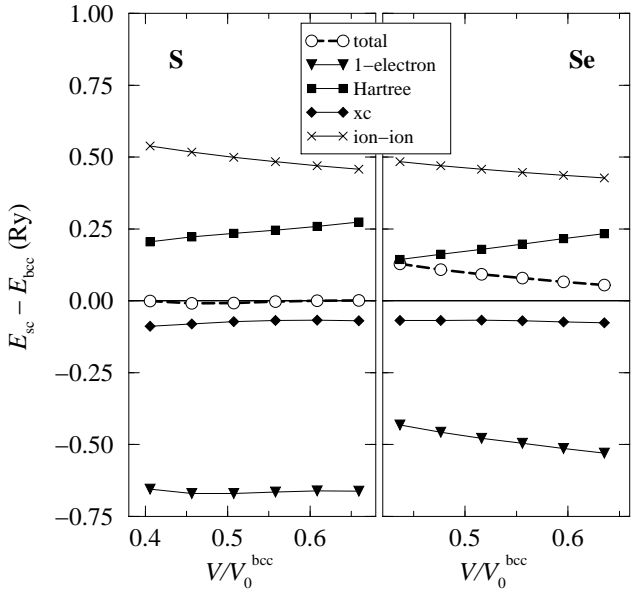,width=3.3in,clip=}}
\caption{
Total energy difference between the simple cubic and bcc
structures as a function of volume for S and Se. Also
shown are the energy differences in the one-electron, Hartree,
exchange-correlation (xc), and ion-ion contributions to the total energy.
Volumes are measured with respect to the theoretical zero-pressure
volumes of $V_0^{bcc}=98.6$ and 125.9 a.u./atom calculated for
bcc S and Se respectively.
}
\label{fig2}
\end{figure}

Figure~\ref{fig2} compares the relative stability of the
sc and bcc structures of S and Se as a function of volume.
To facilitate the comparison, the volumes are measured
with respect to the calculated equilibrium bcc volumes of
$V_0^{bcc}=98.6$ and 125.9 a.u./atom for S and Se, respectively.
The total energy can be divided into
the one-electron contribution,
which arises from the noninteracting kinetic energy and the 
electron--ion interaction, the electron--electron Coulomb
energy,
the exchange and correlation energies, and
the ion--ion energy, which includes the Ewald energy
and a term that accounts for the difference between the pseudopotential 
and the pure Coulomb potential of the ions.
In both materials, the repulsive Hartree and ion--ion
terms favor the more closely packed bcc structure, which has
a more uniform distribution of both ionic and electronic charge.
The attractive electron-ion Coulomb
interaction favors the more open sc structure which tends 
to have very nonuniform charge distributions.
The difference in total energy between the two structures
is smaller than differences in the individual contributions
to the total energy. 
Figure~\ref{fig2} shows that the one-electron contribution
in S favors the sc structure significantly more strongly than
it does in Se, thereby tipping the balance to stabilize
the open sc structure.
Further decomposition of the one-electron contribution shows that
the kinetic energy favors the
more uniform bcc structure. It is thus the electron-ion
interaction that stabilizes the sc S phase. 

The importance of the electron-ion interaction in 
producing different stable structures in S and Se
can be understood in terms of the difference in their
cores. In particular, since the Ne core of S contains only 
$s$ and $p$ electrons, the 3$d$ states in S have no orthogonality constraint
with the core, resulting in a strongly attractive $d$ pseudopotential
in the core region. On the other hand,
in Se, which has $d$ states in the core, the 
repulsive Pauli component of the pseudopotential largely cancels the attractive
Coulomb component in the core region, 
resulting in a relatively weak $d$ potential. 

\begin{figure}
\centerline{\psfig{file=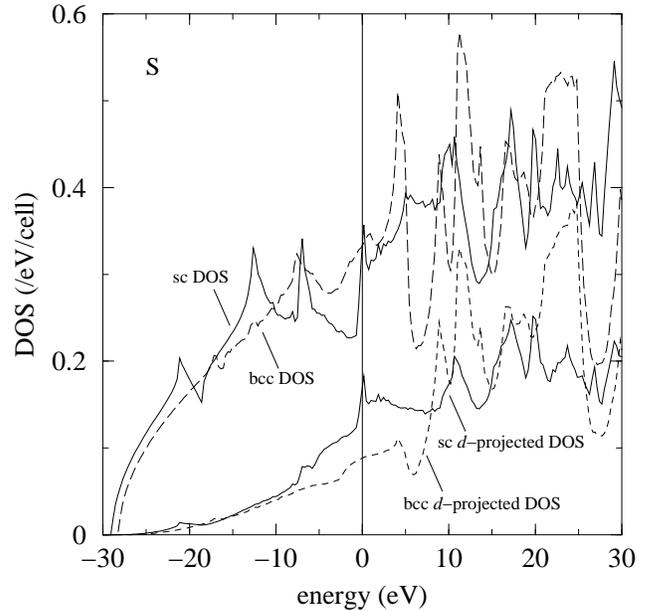,width=3.3in,clip=}}
\caption{
Electronic density of states of bcc and sc S calculated at
a volume of 50 a.u./atom. The Fermi level is at 0 eV.
The $d$-projected density of states,
plotted in lighter lines, shows that the sc structure
has broader $d$-bands and a higher occupation of $d$ states.
}
\label{fig3}
\end{figure}

Under ambient conditions,
the valence $d$ states are unoccupied in both S and Se,
but with compression, the $d$ bands broaden and
eventually cross the Fermi level.
The electronic density of states for bcc and sc S are shown
in Fig.~\ref{fig3}. The volume of 50 a.u./atom
corresponds to about 250 GPa, which is close to the pressure at which 
the sc phase is calculated to become stable. The thin lines
show the $d$-projected
density of states. Since the $d$ bandwidth
varies roughly as $C/d_{\rm nn}^5$, where $C$ is the coordination
number and $d_{\rm nn}$ is the nearest neighbor distance,\cite{harrison} 
the shorter bonds 
in the sc structure more than compensate for the lower coordination, 
leading to broader $d$ bands in the sc phase as 
compared to the bcc phase. With more $s$-$d$ transfer of
electrons in the sc phase than in the bcc (or fcc) structures,
the deep S $d$ pseudopotential becomes important in the 
energetics, stabilizing the sc phase. In Se, the sc
phase similarly has larger $d$ occupation than the fcc and bcc phases, 
but because the $d$ potential is weak, this $s$-$d$ transfer does not 
lower the one-electron energy enough to stabilize the sc structure. 

\begin{figure}
\centerline{\psfig{file=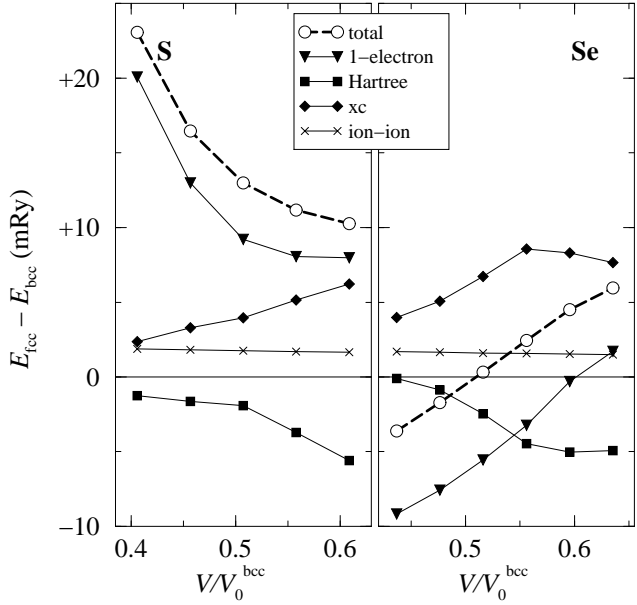,width=3.3in,clip=}}
\caption{
Total energy difference between the fcc and bcc
structures as a function of volume for S and Se. Also
shown are the energy differences in the one-electron, Hartree,
exchange-correlation (xc), and ion-ion contributions to the total energy.
Volumes are measured with respect to the theoretical zero-pressure
volumes of $V_0^{bcc}=98.6$ and 125.9 a.u./atom calculated for
bcc S and Se respectively.
}
\label{fig4}
\end{figure}

Similar reasoning helps explain the absence
of the close-packed fcc structure in high-pressure S.
Fig.~\ref{fig4} compares the relative stability
of the fcc and bcc phases of S and Se as a function of volume.
Again the different structural preferences 
of S and Se can be attributed to differences in the one-electron
contributions to the total energy. With increasing compression,
the one-electron energy in S increasingly favors the bcc structure,
while in Se it increasingly favors the fcc structure.
In both materials, the kinetic energy favors the 
more uniformly distributed fcc structure while the electron--ion term
favors the bcc structure. With 
the deeper $d$ pseudopotential in S,
the importance of the electron-ion term in S is enhanced, 
stabilizing the bcc structure.
The difference in openness between the fcc and bcc structures is of course
much smaller than that between the sc and bcc structures, so in
comparisons of the fcc and bcc structures, the energy
differences are smaller and the energy balance is more subtle.

This discussion of the differences between S and Se and the
important role of the presence of $d$ states in the core suggests
that the heavier chalcogen Te should be similar to Se in terms
of high-pressure structural phases. Indeed, like Se, Te is observed to 
transform from the $\beta$-Po structure to the bcc structure with
pressure.\cite{parthasarathy88} In our calculations, as well as earlier ones by
Kirchhoff et al.,\cite{kirchhoff94} this transition in Te is found close to the 
experimentally measured transition pressure
of 27 GPa. In addition, we find that the close-packed fcc structure 
becomes stable above about 80 GPa.  
Finally, as in Se, the sc structure is calculated to be energetically
unfavorable as a high-pressure phase of Te.\cite{kirchhoff94}

\section{PHONON SPECTRA, ELECTRON-PHONON COUPLING, AND SUPERCONDUCTIVITY}

\subsection {Method}
The density-functional linear response method\cite{baroni87}
 is used to compute the
phonon spectra for each structure at several different pressures. 
The dynamical matrices are computed using the scheme
described in Ref. \onlinecite{quong92}, where the self-consistent change
in the Hamiltonian caused by ionic displacements is
obtained by solving a Bethe-Saltpeter equation for the
change in the charge density.
The dynamical matrix is computed on a grid of $6^3$ phonon
wavevectors ${\bf q}$ for $\beta$-Po S and $8^3$ wavevectors for sc
and bcc S.  For Se, $8^3$, $10^3$, and $12^3$ ${\bf q}$ points are 
sampled in the full $\beta$-Po, bcc, and fcc Brillouin zones, respectively. 

The matrix element for scattering of an electron from state $n{\bf k}$
to state $n^\prime {\bf k}^\prime$
by a phonon with frequency $\omega_{{\bf q}\nu}$ and eigenvector
${\bf \hat{\epsilon}}_{{\bf q}\nu}$ is given by
\begin{equation}
g(n{\bf k},n^\prime{\bf k}^\prime,{\bf q}\nu)=
  \sqrt{\hbar \over 2M\omega_{{\bf q}\nu} }
       \langle n{\bf k}| {\bf \hat{\epsilon}}_{{\bf q}\nu} \cdot 
               {\bf \nabla_R} V_{\rm SCF}
                       | n^\prime {\bf k}^\prime \rangle,
\end{equation}
where ${\bf \nabla_R} V_{\rm SCF}$ is the gradient of the 
self-consistent potential with respect to atomic displacements. 
This scattering gives rise to a finite
phonon linewidth,
\begin{equation}
\gamma_{{\bf q}\nu} = 2 \pi \omega_{{\bf q}\nu} [N(E_F)]^2 \langle\langle |g_{{\bf q}\nu}|^2 \rangle\rangle,
\end{equation}
where $N(E_F)$ is the electronic density of states per spin at the
Fermi level.  The double brackets $\langle\langle ... \rangle\rangle$
denote a doubly constrained
Fermi surface average as defined in Ref. \onlinecite{lam86}.
This scattering process also contributes to the effective mass of
the electrons via the mass enhancement parameter
$\lambda$, obtained from the wavevector- and branch-dependent contributions
\begin{equation}
\lambda_{{\bf q}\nu}
= {\gamma_{{\bf q}\nu} \over \hbar \pi\ N(E_F) \omega_{{\bf q}\nu}^2} 
\end{equation}
 by summing over branches $\nu$ and averaging over wavevectors ${\bf q}$.

The electron-phonon spectral function, which
measures the effectiveness of phonons of a given energy
to scatter electrons on the Fermi surface, plays a central role
in the Eliashberg strong-coupling theory of superconductivity.
The spectral function is
given by
\begin{equation}
\alpha^2F(\omega) = {1 \over 2 \pi N(E_F)} \sum_{{\bf q}\nu}
    \delta(\omega-\omega_{{\bf q}\nu})
     {\gamma_{{\bf q}\nu} \over \hbar \omega_{{\bf q}\nu}}.
  \end{equation}
Within this framework, the mass enhancement parameter $\lambda$ is
proportional to the inverse-frequency moment of the spectral function.

In the density-functional linear-response method, 
${\bf \nabla_R} V_{\rm SCF}$ is
computed in the process of determining the dynamical
matrices.   Therefore we calculate the electron-phonon
matrix elements on the same grid of 
wavevectors used in the phonon calculations. 
For calculations of the
phonon density of states and electron-phonon spectral function,
which involve summing over {\bf q} points throughout the Brillouin
zone, a Fourier interpolation procedure is used
to obtain the dynamical matrix and its dissipative
part on a denser mesh of {\bf q} points.\cite{liu00}
The doubly constrained Fermi surface average
of $g$ is computed
using Bloch functions on dense meshes of
at least $30^3$ {\bf k} and {\bf k$^\prime$} points in the full Brillouin zone, 
with delta functions at the Fermi level replaced by
Gaussians of width of order 0.01 Ry chosen to
reproduce the value of $N(E_F)$ obtained using the
linear tetrahedron method.\cite{lehmann72}

\subsection{Results and discussion}

Tables~\ref{tabS} and \ref{tabSe} 
list parameters that characterize the phonon spectra and
electron-phonon couplings in the high-pressure phases of S and Se. 
Included are the average phonon frequency $\langle\omega\rangle$
and the logarithmic $\lambda$-weighted average phonon frequency
$\langle\omega_{\ln}\rangle = \exp[
       \sum_{{\bf q}\nu} \lambda_{{\bf q}\nu} \ln(\omega_{{\bf q}\nu}) /
                               \sum_{{\bf q}\nu} \lambda_{{\bf q}\nu}]$.
Some of the values for $N(E_F)$ and $\lambda$ listed in Table \ref{tabS}
differ by a few percent from those published in our earlier paper\cite{rudin99}
on S because we use the more accurate tetrahedron method
for calculating the electronic density of states in the present work.  

\begin{table}
\caption{Calculated values of structural, 
vibrational, and electron-phonon parameters for
$\beta$-Po, sc, and bcc S at various pressures. The electronic
density of states at the Fermi level $N(E_F)$ is in units of
states/eV/spin/cell.}
\begin{tabular}{cccccccc}
 & $P$ & $V$ & $\alpha_r$ & $N(E_F)$ &
 $\langle\omega\rangle$ &
 $\langle\omega_{\ln}\rangle$ & $\lambda$ \\
 & (GPa) & (a.u.) & & & (meV) & (meV) & \\
\tableline
 & 160 & 57.87 & 104.0$^\circ$ & 0.148 & 46.4 & 37.7 & 0.76 \\
$\beta$-Po & 200 & 54.22 & 104.1$^\circ$ & 0.143 & 50.6 & 40.2 & 0.76 \\
 & 280 & 48.77 & 104.4$^\circ$ & 0.140 & 57.2 & 42.3 & 0.61 \\
\tableline
 & 280 & 47.80 & 90.0$^\circ$ & 0.159 & 71.8 & 48.1 & 0.52 \\
sc & 320 & 46.14 & 90.0$^\circ$ & 0.146 & 76.5 & 52.8 & 0.44 \\
 & 550 & 39.77 & 90.0$^\circ$ & 0.118 & 91.8 & 57.8 & 0.35 \\
\tableline
bcc & 550 & 38.81 & 109.5$^\circ$ & 0.136 & 67.0 & 50.0 & 0.70 \\
\end{tabular}
\label{tabS}
\end{table}

\begin{table}
\caption{Calculated values of structural, vibrational, and 
electron-phonon parameters for
$\beta$-Po, bcc, and fcc Se at various pressures. The electronic density 
of states at the Fermi level $N(E_F)$ is in units of
states/eV/spin/cell. }
\begin{tabular}{cccccccc}
 & $P$ & $V$ & $\alpha_r$ & $ N(E_F)$ & 
 $\langle\omega\rangle$ & $\langle\omega_{\ln}\rangle$ & $\lambda$ \\
 & (GPa) & (a.u.) & & & (meV) & (meV) & \\
\tableline
 & 60 & 93.33 & 104.1$^\circ$ &  0.180 & 23.5 &  18.8 & 0.58 \\
$\beta$-Po & 80 & 87.37 & 104.8$^\circ$ &  0.176 & 25.5 &  19.5 & 0.54 \\
 & 100 & 83.20 & 105.2$^\circ$ &  0.174 & 27.4 &  21.6 & 0.50 \\
\tableline
   & 120 & 78.73 & 109.5$^\circ$ & 0.201 & 26.6 & 17.6 & 0.89  \\
   & 140 & 75.80 & 109.5$^\circ$ & 0.193 & 28.3 & 18.8 & 0.76  \\
bcc & 160 & 73.32 & 109.5$^\circ$ & 0.187 & 29.4 & 17.6 & 0.77  \\
   & 220 & 67.36 & 109.5$^\circ$ & 0.182 & 33.3 & 21.2  & 0.73  \\
   & 260 & 64.25 & 109.5$^\circ$ & 0.175 & 35.6 & 21.4 & 0.71 \\ 
\tableline
fcc & 260 & 63.82 & 60.0$^\circ$ & 0.183 & 34.93 & 25.23 & 0.59 \\
\end{tabular}
\label{tabSe}
\end{table}

Within each structural phase,
compression stiffens the lattice and weakens the
electron-phonon interaction, as indicated 
by the rise in $\langle\omega\rangle$ and
fall in $\lambda$. The pressure-dependence of $\lambda$
varies from structure to structure.  In sc S, 
$\lambda$ varies approximately as $1/\langle\omega\rangle^2$, as might be
expected from Eq. (3).  In $\beta$-Po S, $\beta$-Po Se, and bcc Se, however,
the pressure-dependence of $\lambda$ is much weaker.
These differences can be understood by examining contributions
to $\lambda$ from different phonon modes.

\begin{figure}
\centerline{\psfig{file=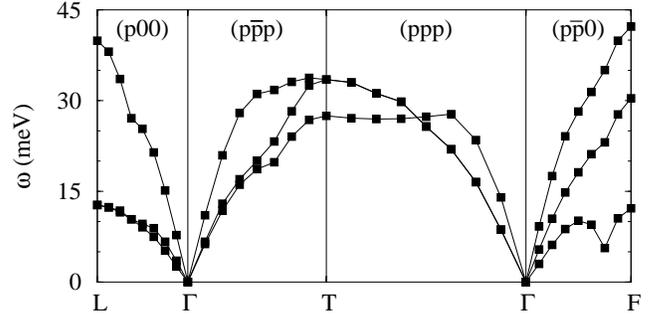,width=3.3in,clip=}}
\caption{ Phonon dispersion curves in $\beta$-Po Se at
60 GPa. Directions are specified in terms of 
reciprocal-lattice basis vectors. Soft phonon anomalies related
to nesting of the Fermi surface are evident along
the $\Gamma$ to F line. 
}
\label{fig5}
\end{figure}

\begin{figure}
\centerline{\psfig{file=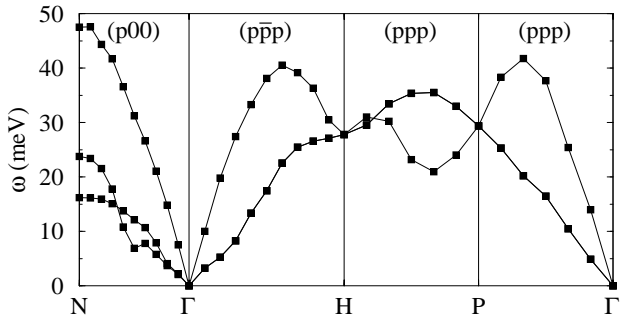,width=3.3in,clip=}}
\caption{ Phonon dispersion curves in bcc Se
at 120 GPa. Directions and polarizations are specified in terms of
reciprocal-lattice basis vectors. Phonon anomalies are present in the
($\bar{1}$11)-polarized branch along $\Gamma$ to N, and
in the transverse branches along $\Gamma$ to H.
 }
\label{fig6}
\end{figure}

In some of the structural phases, it is possible to identify 
specific phonon modes that contribute strongly to $\lambda$.  
In the $\beta$-Po phase of Se, for example, there are strong 
anomalies in the phonon dispersion curves along the $\Gamma$ to F direction
({\it i.e.}, the $(1\bar{1}0)$ direction in terms of the
reciprocal lattice vectors ${\bf b}_i$),
as shown in Fig.~\ref{fig5}. Similar but less pronounced anomalies exist
in the $\beta$-Po phase of S. These anomalies are associated
with nesting of the Fermi surface by the wavevector ${\bf q} \approx
0.7({\bf b}_1-{\bf b}_2)$, as 
confirmed by calculations of the geometric nesting factor 
$\langle\langle\delta({\bf k}-{\bf k^\prime}-{\bf q})\rangle\rangle$.
The strong Fermi surface nesting affects not only the phonon frequencies,  
but the linewidths as well. The combination of low frequencies
and enhanced linewidths results in a large contribution to
$\lambda$ from this part of the $\beta$-Po Brillouin zone.   
Anomalies are also present in the phonon spectra of bcc S and bcc Se.
Figure~\ref{fig6} shows the dispersion curves for bcc Se at 120 GPa.
While not correlated with nesting vectors,
the low-frequency anomalies in the transverse branches along 
the $\Gamma$ to N and $\Gamma$ to H lines nevertheless give large 
contributions to $\lambda$. 

The soft phonon anomalies in the bcc and $\beta$-Po phases 
are more weakly pressure-dependent than the overall phonon
spectrum. The pressure dependence of $\lambda$ in these phases is strongly
influenced by these persistent anomalies and is therefore weaker than 
the $\langle\omega\rangle^{-2}$ dependence suggested by Eq. (3). 
In sc S, the electron-phonon coupling is not found to be dominated
by any particular phonon mode, 
so the pressure dependence of $\lambda$ is to a good approximation 
determined by that of the average phonon frequency. 

\begin{figure}
\centerline{\psfig{file=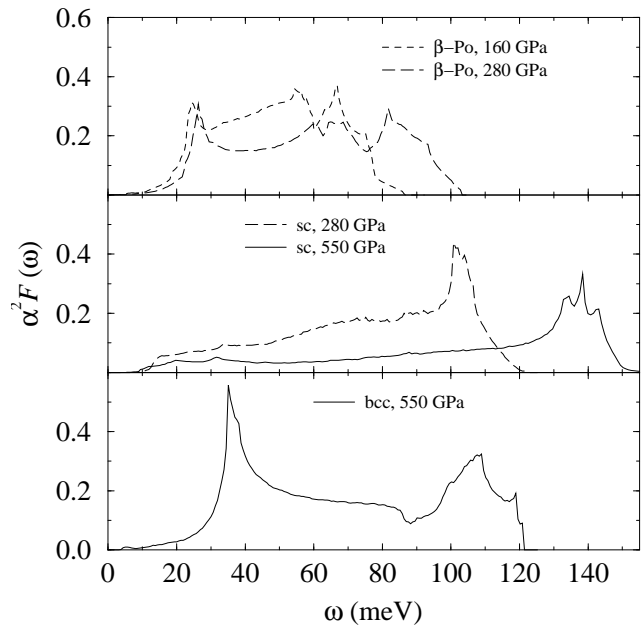,width=3.3in,clip=}}
\caption{ Electron-phonon spectral functions calculated for $\beta$-Po,
sc, and bcc S. 
}
\label{fig7}
\end{figure}

As can be seen in Tables~\ref{tabS} and \ref{tabSe}, 
transformations from structure to structure 
are usually accompanied by abrupt changes in the phonon and electron-phonon
parameters. In comparing the various structural phases of S, the simple-cubic 
phase stands out in having the weakest electron-phonon coupling parameters. 
The spectral function for sc S, plotted in Fig.\ref{fig7},  
is strikingly different from those of bcc and $\beta$-Po S. 
Both the average and maximum phonon frequencies
in the sc phase are much larger than those
in the other phases.
The higher frequencies in the sc 
phase are due to the shorter nearest-neighbor 
distances and the strong concentration of charge
along the bonds in the open sc lattice. 
With these stiff bonds, the $(1\bar{1}0)$-polarized transverse
branch along the (110) direction and both of the transverse
branches along the (111) direction are nearly as high in energy as the 
longitudinal branches. This leads to a phonon density of states and 
a spectral function that
are dominated by a high-frequency peak.   
The large phonon energy scale and
the lack of low-energy modes that couple strongly to electrons, as 
evidenced by the relatively large value of $\langle \omega_{\ln} \rangle$,
combine to drive $\lambda$ down in the sc phase of S. 

\begin{figure}
\centerline{\psfig{file=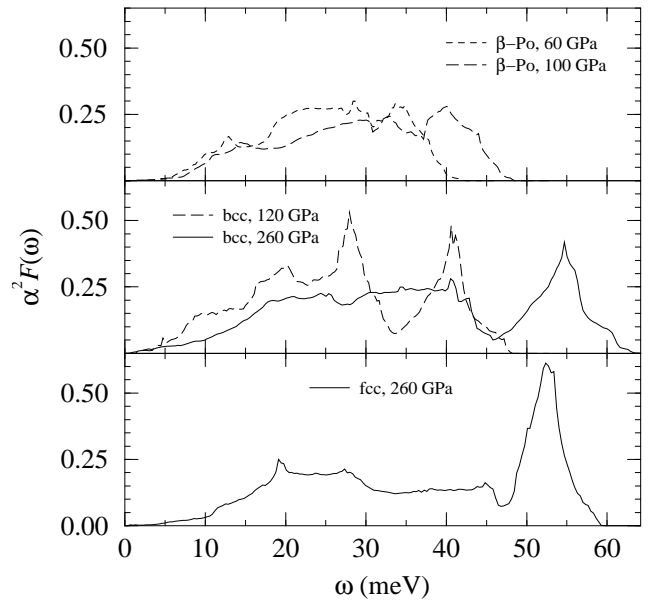,width=3.3in,clip=}}
\caption{ Electron-phonon spectral functions calculated for $\beta$-Po,
bcc, and fcc Se. 
}
\label{fig8}
\end{figure}

Among the high-pressure phases of Se, the bcc phase stands out
in that $\lambda$ is largest in this phase. Although the average
phonon frequencies are similar to those in the neighboring 
phases near the transition pressures, 
$\langle \omega_{\ln} \rangle$ is significantly depressed in the bcc phase.
This is a sign of the importance of the low-frequency
anomalies already identified in the bcc Se phonon spectra.  
These anomalies enhance the spectral function at low energies,
as shown in Fig.\ref{fig8}, 
and they have a significant effect on $\lambda$ because they
appear at low frequencies.
Although $\beta$-Po Se also has soft phonon anomalies,
there are fewer of them due to the lower symmetry of the lattice,
and they have a smaller effect on $\lambda$.   
No comparable soft phonon anomalies are found in the fcc phase.

In comparing analogous structural phases of S and Se, the 
differences in electron-phonon coupling
strength reflect not only differences in the phonon spectra, but also
differences in the electron-phonon matrix elements.    
The stronger potential in S gives rise to a larger change in the bare
potential 
${\bf \nabla_R} V_{\rm bare}$ 
when atoms are displaced. Furthermore,
because the $d$ electrons are more tightly bound in S, these electrons
are less effective in screening 
${\bf \nabla_R} V_{\rm bare}$.
In combination, these two effects result in significantly
larger matrix elements of 
${\bf \nabla_R} V_{\rm SCF}$ 
in S than in Se.
Combined with the smaller mass in S, this leads to larger
linewidths. This is evident in both the $\beta$-Po 
and bcc phases, where the phonon frequencies
in S are a factor of two or more higher than those in Se, yet
the mass enhancement parameters in the two materials are 
close. 

Our results can be compared to earlier density-functional-based studies of 
the electron-phonon interaction in $\beta$-Po and 
bcc Te and bcc S.\cite{zakharov95,mauri96}
Soft phonon anomalies very similar to those discussed 
here have been identified to be important 
in the electron-phonon coupling in $\beta$-Po and bcc Te.\cite{mauri96}
In addition, the electron-phonon mass enhancement 
parameter has been calculated to undergo a large jump at the $\beta$-Po to bcc 
transition in Te, just as we find for Se. 
For bcc S, our value of $\lambda = 0.70$ at 550 GPa is somewhat larger
than the value of 0.58 reported in Ref. \onlinecite{zakharov95} for 
a slightly higher pressure of 584 GPa. Since
values of $\lambda_{{\bf q}\nu}$ are in reasonable agreement
along high-symmetry directions, it is likely that the difference arises
from a difference in sampling of wavevectors:
In the present work, a uniform grid of points throughout the Brillouin zone
is sampled, while in Ref. \onlinecite{zakharov95},
$\lambda_{{\bf q}\nu}$ is calculated along a few high-symmetry directions and
spherically averaged to estimate $\lambda$.

\subsection{Making contact with experiments}

The need to carry out experiments inside
high-pressure cells limits the types of
experimental probes available to investigate the electron-phonon interaction 
in the metallic phases of the chalcogens.  In particular,
the most common probe of $\alpha^2F$ in
superconductors, quasiparticle tunneling,\cite{wolf85} 
is challenging 
because of the need to make well characterized tunnel junctions
connected to leads that enter the high-pressure cell.\cite{wright77}

Measurement of superconducting properties
such as
the transition temperature $T_C$, the thermodynamic
critical field $H_C$, the zero-temperature gap $\Delta_0$,
and the isotope effect on $T_C$, would be
one avenue for probing the electron-phonon interaction.\cite{carbotte90}
Since superconducting properties depend not only on the 
electron-phonon spectral function $\alpha^2F$, but also on the Coulomb 
pseudopotential $\mu^*$, it is necessary to measure a combination of 
superconducting properties to get a handle 
on both the electron-electron and the electron-phonon parameters. 
In the weak coupling regime, $T_C$, $\Delta_0$, and
$H_C$ all essentially depend on the difference between
$\lambda$ and $\mu^*$, rather than on $\alpha^2F$ and $\mu^*$ 
separately. In this case, $T_C$ and the isotope effect
on $T_C$, with their different $\alpha^2F$ and $\mu^*$ dependences, 
would be the most useful combination of measurements. 

\begin{figure}
\centerline{\psfig{file=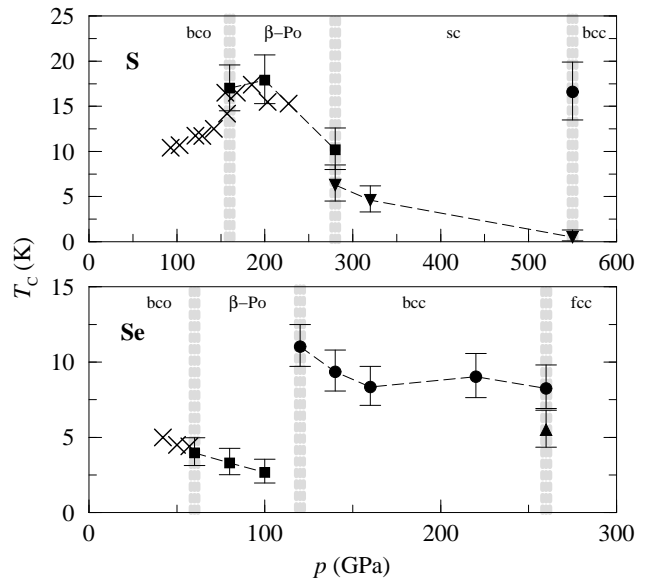,width=3.3in,clip=}}
\caption{ Superconducting transition temperature of S and Se as a function
of pressure. Squares, circles, up triangles, and down triangles
represent results of calculations for the $\beta$-Po, bcc, fcc, and sc
structures, respectively.  The error bars show the range of
$T_C$ for $0.09<\mu^*(\omega_{\rm max})<0.13$
and the data points correspond to $\mu^*(\omega_{\rm max})=0.11$. 
Experimental data are plotted as crosses. 
The dashed lines connecting calculated points
serve as guides to the eye, and the vertical gray
bars roughly separate the pressure ranges over which 
different structures are stable.
}
\label{fig9}
\end{figure}

Figure 9 shows our calculated superconducting transition temperatures
for the high-pressure phases
of S and Se for different values of $\mu^*$,
obtained by using the calculated $\alpha^2F$ functions
as input to the Eliashberg equations.
Available experimental data 
for bco Se,\cite{akahama92} bco S,\cite{struzhkin97} and
$\beta$-Po S,\cite{struzhkin97,gregoryanz00} 
are included. 
The theoretical data points in Fig.~\ref{fig9}
correspond to $\mu^*(\omega_{\rm max})=0.11$, where $\omega_{\rm max}$ is
the maximum phonon frequency.
This representative value of $\mu^*(\omega_{\rm max})$ is chosen
because it reproduces the experimentally 
measured $T_C$ of 17 K at 160 GPa in S.\cite{struzhkin97} 
The error bars show the effect of varying
$\mu^*(\omega_{\rm max})$ by $\pm 0.02$.
While there is no reason to assume $\mu^*$
remains constant with changes in pressure or structure,
or that $\mu^*$ in Se should be the same as in S, $\mu^*$ typically
lies within the range of 0.1 to 0.14 for a wide range of materials. 
Furthermore, a rough upper bound for $\mu^*$ given by
$[\ln(E_F/\omega_{\rm max})]^{-1}$ suggests $\mu^*$ is less than
about 0.16 in all the phases considered here. 

The calculated transition temperatures 
follow the same trends as $\lambda$, with large increases in $T_C$ upon
transition to both bcc Se and bcc S, and downward
jumps in $T_C$ at the bcc$\rightarrow$fcc transition in Se and the
$\beta$-Po$\rightarrow$sc transition in S.  
As reported earlier,\cite{rudin99}
our analysis attributes
the large $T_C$ of 17 K in $\beta$-Po S not so much to 
strong electron-phonon coupling, but rather to a combination
of moderate coupling and a large phonon energy scale. 
For analogous structural phases in S and Se, the primary
reason for the higher transition temperatures in S is the
stiffer lattice, which sets the energy scale for $T_C$.

The high phonon frequencies suggest that vertex corrections might be
important in these materials.
The other determining factor for the strength of vertex corrections
is the magnitude of the electron-phonon interaction,\cite{rudin98}
which is moderate in both S and Se.
Indeed, we estimate that the effect of the vertex
corrections on $T_C$ is on the order of one percent,
and can be neglected given the uncertainty in $\mu^*$.

Although we have used the
Eliashberg formalism to calculate $T_C$ for all the phases, 
the coupling is sufficiently weak that the 
Allen-Dynes approximate formula for $T_C$\cite{allen75} gives similar results. 
Furthermore, the ratios 
$2\Delta_0/k_BT_C$ and $\gamma T_C^2/H_C^2$, where $\gamma$
is the linear coefficient of the specific heat,
are calculated to be close to the BCS values
of 3.53 and 1.68, respectively. For $\beta$-Po S at 160 GPa,
for example, we find $2\Delta_0/k_BT_C$ = 3.70 and $\gamma T_C^2/H_C^2$
=1.66, assuming $\mu^*(\omega_{\rm max})$ = 0.11. 

As suggested above, measurement of the isotope effect
on $T_C$ would be one way to obtain more information on
the Coulomb parameter. For S at 160 GPa, we estimate that
the isotope exponent $\alpha=-d\ln{T_C}/d\ln{M}$
changes from 0.48 to 0.43 as $\mu^*$ is varied from 0.09 to 0.13,
assuming isotopic masses $M$ of 32 and 36. To distinguish
between $\mu^*$ = 0.09 and $\mu^*$ = 0.13 then requires 
being able to measure $T_C$ to an accuracy of better than 0.1 K.

\begin{figure}
\centerline{\psfig{file=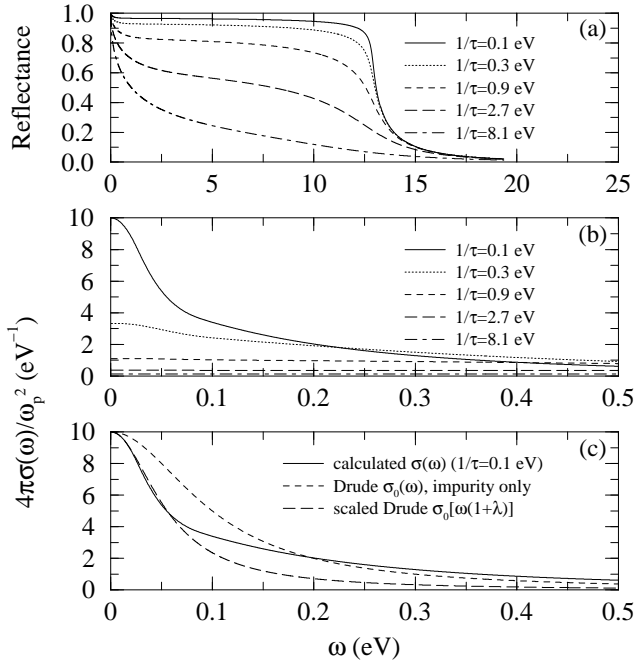,width=3.3in,clip=}}
\caption{ Normal-state (a) reflectance and (b) optical conductivity 
calculated for $\beta$-Po S at 160 GPa and T=0 K. 
Results are shown for different impurity scattering
times $\tau$. In (c), the optical conductivity
with $\tau$ = 0.1 eV is redrawn along with the
corresponding Drude conductivity $\sigma_0(\omega)$ assuming scattering by 
impurities only. Also plotted is $\sigma_{0}[\omega(1+\lambda)]$,
showing that electron-phonon scattering narrows
the Drude peak by a factor of $1+\lambda$. 
}
\label{fig10}
\end{figure}

Infrared spectroscopy can be used to 
probe $\alpha^2F$ independent of $\mu^*$.\cite{allen71,marsiglio97}
In the normal state, the phonons contribute to the  optical conductivity
via their renormalization of the electronic quasiparticles and by
phonon-assisted scattering processes.  
Panels (a) and (b) in Fig.~\ref{fig10} show
the normal-state $T=0$ reflectance and optical conductivity 
calculated for different values of impurity scattering
rates for $\beta$-Po S at 160 GPa. The expected Holstein 
structure, which occurs at frequencies near the phonon frequencies, 
is small because the coupling strength is relatively weak.
Hence, it is likely that the extraction of $\alpha^2F$ from 
phonon structure in the optical conductivity data would be 
challenging for this system. However, optical conductivity
data could still be useful for estimating $\lambda$. 
The electron-phonon interaction
narrows the Drude peak associated with impurity scattering
by a factor of 1+$\lambda$
due to the renormalization of electronic quasiparticle energies and
weights and transfers the low-frequency weight to the mid-infrared region,
as illustrated in Fig.~\ref{fig10}(c).
With an independent determination of the plasma
frequency $\omega_p$, the width and height of the Drude peak
could be used to estimate $\lambda$. 

Perhaps a simpler way to determine $\lambda$ from
experiments is via the dc conductivity. 
At temperatures on the order of the Debye temperature,
the phonon-limited electrical resistivity is linear in temperature
with a slope proportional to $\lambda_{\rm tr} =
2\int {\rm d}\omega \alpha_{\rm tr}^2F(\omega)/\omega$.
Like $\alpha^2F$, the transport spectral function $\alpha_{\rm tr}^2F$ 
measures the effectiveness
of phonons to scatter electrons on the Fermi surface,
but it is weighted to take into account the
change in direction of the electron velocity.\cite{allen71}
Experimental measurements of the electrical resistivity have been reported
for the bco and lower pressure phases of S and Se.\cite{bundy80,akahama92}
Extensions of these measurements to higher pressures
would test our results for the electron-phonon coupling strength.
Of particular interest is the linear coefficient of the resistivity
in $\beta$-Po S. We calculate $\lambda_{\rm tr} = 0.78$ at
160 GPa, close to our result for $\lambda$.
We believe this is the most direct way to verify
our theoretical prediction that,
despite the large $T_C$, the electron-phonon
coupling is not particularly strong in this phase.
  
\section{CONCLUSIONS}

We have presented density-functional based calculations
on the pressure-induced structural phase transitions
in S and Se.   
Although there are some similarities in
the high-pressure phase diagrams of these isovalent elements,
there are also striking differences. With increasing 
compression, Se adopts a sequence of ever more closely packed 
structures ($\beta$-Po$\rightarrow$bcc$\rightarrow$fcc), while S favors 
more open structures ($\beta$-Po$\rightarrow$sc$\rightarrow$bcc).
These differences can be understood in terms
of the deeper $d$ pseudopotential in S arising from the
lack of $d$ states in the S core.

All the high-pressure phases of S and Se are
calculated to have moderate electron-phonon coupling
strengths and superconducting transition temperatures 
in the range of 0.5-20 K.
Structure and pressure dependences of the
coupling strength and $T_C$ can be
understood in terms of changes in the phonon spectra.
In particular, the large observed $T_C$ in $\beta$-Po S is 
calculated to arise from a combination of a moderate 
mass enhancement parameter $\lambda$ and a large overall phonon 
energy scale. The large increase
in $\lambda$ and $T_C$ predicted upon transformation to the bcc
phase in both S and Se is due to the presence of
soft modes in the bcc phonon spectra that
strongly couple to electrons.

We suggest that measurements of the electrical resistivity
as a function of temperature, the transition temperature $T_C$, and 
the isotope effect on $T_C$ would be the most promising avenues 
for testing our
results on the electron-phonon interaction in these
compressed phases. Further, $T_C$ measurements could be used to search
for the predicted sc phase of S and the predicted fcc phase of Se
since our results indicate that $T_C$ should change abruptly 
when these structural transformations take place.  The calculated
transition pressures for both sc S and fcc Se are within
the range of current diamond-anvil-cell experiments. 

\section{ACKNOWLEDGMENTS}  %%%%%%%%%%%%%%%%%%%%%%%%%%%%%%%%%%%%

We thank R.~J.~Hemley and V.~V.~Struzhkin
for valuable discussions.
This work was supported by the National Science Foundation under Grant
DMR-9973225 and by the NPACI.


\begin{references}
\bibitem{bcofootnote} The bco structure, 
which is observed experimentally in S, Se, and Te just below the $\beta$-Po
structure in pressure, is not considered in this work
because we find it to be not only higher in energy than,
but also mechanically unstable with respect to the $\beta$-Po 
structure. The instability is found in both the local-density
and generalized-gradient approximations. Similar results for
the bco structure are reported in Refs. 11 and 18.
\end{references}
\end{document}